\documentclass[a4paper,11pt]{article}
\pdfoutput=1 
\usepackage{jheppub} 
\usepackage[T1]{fontenc}
\usepackage{amsmath}
\usepackage{slashed}
\newcommand{\be}{\begin{equation}}
\newcommand{\ee}{\end{equation}}

\title{Increase of the Energy Necessary to Probe Ultraviolet Theories Due to the Presence of a Strong Magnetic Field}

\author[a]{Rodolfo P. Mart\'{\i}nez-y-Romero,}
\author[a]{Leonardo Pati\~no,}
\author[a,1]{Tiber Ram\'{\i}rez-Urrutia.\note{Also at Escuela de Bachilleres, Instituto de Educaci\'on Media Superior del Distrito Federal Plantel Milpa Alta.}}

\affiliation[a]{Departamento de F\'{\i}sica, Facultad de Ciencias, Universidad Nacional Aut\'noma de M\'exico,\\Apartado Postal 70-542, CP 04510, Mexico City, Mexico}

\emailAdd{rmr@ciencias.unam.mx}
\emailAdd{leopj@ciencias.unam.mx}
\emailAdd{tiber@ciencias.unam.mx}

\abstract{We use the gauge gravity correspondence to study the renormalization group flow of a double trace fermionic operator in a quark-gluon plasma subject to the influence of a strong magnetic field and compare it with the results for the case at zero temperature and no magnetic field, where the flow between two fixed points is observed. Our results show that the energy necessary to access the physics of the ultraviolet theory increases with the intensity of the magnetic field under which the processes happen. We provide arguments to support that this increase is scheme independent, and to exhibit further evidence we do a very simple calculation showing that the dimensional reduction expected in the gauge theory in this scenario is effective up to an energy scale that grows with the strength of such a background field. We also show that independently of the renormalization scheme, the coupling of the double trace operators in the ultraviolet fixed point increases with the intensity of the background field. These effects combined can change both, the processes that are expected to be involved in a collision experiment at a given energy and the azimuthal anisotropy of the measurements resulting of them.}

\begin{document} 
\maketitle
\flushbottom

\section{Introduction}
\label{sec:intro}

The holographic renormalization program has been part of the gauge/gravity correspondence almost since its origins \cite{Bianchi:2001de, Bianchi:2001kw, Skenderis:1999nb, Freedman:1999gp, deBoer:2000cz}, and in particular the Wilsonian approach within this program has been the object of much attention in recent years \cite{Heemskerk:2010hk,Faulkner:2010jy,deBoer:1999tgo,deBoer:2000cz,Skenderis:2002wp}. This approach provides a systematic framework to treat the properties of the renormalization flow in a gauge theory at a non-perturbative level by means of calculations in a dual gravitational theory \cite{Skenderis:2002wp, Balasubramanian:2012hb}.

The staring point is the AdS/CFT correspondence which provides a duality between type IIB string theory in AdS$^5\times$S$_5$ and a {\cal N}=4 supersymmetric conformal $SU(N)$ field theory in a four dimensional Minkowski space. This original correspondence has been modified in a number of ways, for instance by allowing the AdS$^5$ part of the geometry to have an horizon, becoming an AdS-Schwarzschild solution, which corresponds to the introduction of a finite temperature in the gauge theory. A number of modifications of the background metric have been introduced to model different physical scenarios, and in general these modifications are admissible as long as a five dimensional subspace still approaches asymptotically AdS$^5$ space close to its boundary, and the remaining compact subspace retains enough symmetry to describe the dual gauge theory. In all these constructions, the directions along the compact manifold are dual to internal degrees of freedom in the gauge theory, while the directions along the boundary of the asymptotic AdS$^5$ are in correspondence with the directions in which the dual theory propagates. Of particular importance to the renormalization program is that the direction that extends away from the boundary into the bulk of the asymptotic AdS$^5$ space, that is, the radial direction of this space, is related to the energy scale in the gauge theory, which is also a distance scaling.

Since the gauge/gravity correspondence relates the high energy behavior of the field theory with the low energy regime of the string theory, to study high energy processes in the field theory, we can approximate its dual to be governed by the low energy limit of type IIB string theory, that is, type IIB supergravity, making it possible to work with a classical action which exponential is the approximated dual to the quantum generating functional \cite{Heemskerk:2010hk,deBoer:1999tgo,deBoer:2000cz,Skenderis:2002wp}.

Operators in the field theory side are related to fields in the gravity component of the correspondence, so to study the behavior of an operator in the gauge theory we must find its dual field. One general approach to finding this field is by determining the one that matches the symmetries of the operator. It will be important in this work to notice that given the correspondence between the radial direction in the gravity side and the energy scale in the field theory side, a normalizable field that close to the boundary behaves like $m r^{-\Delta}$ is dual to an operator of scaling dimension $\Delta$ and expectation value $m$.

Two theories being dual implies that all the degrees of freedom of one of them have to be codified in the other and vice versa. Holografic renormalization relays on the fact that this should also be true for the renormalization of the fields in the gauge theory, and looks for the way in which this information is encoded in the gravitational side of the duality. In particular, the encoding of the Wilsonian approach to renormalization can be found by, instead of extracting the information of the field theory from a precisely radially localized four dimensional surface in the bulk, integrating out a fraction of the space very near to the boundary and writing it as a surface term on the limiting surface of the integration domain. This integration, being on the radial direction extending away from the boundary and lying very close to it, corresponds to integrating out the high energy modes in the gauge theory, as it should be in the Wilsonian approach to renormalization.

More explicitly, what is needed is to write the gravity action as an integral over the bulk up to a surface close to the boundary, plus the corresponding surface integral just discussed, in such a way that the total action is independent of the location of this surface. This requirement can be expressed as a radial Hamilton-Jacobi equation, where, as the limiting surface is moved, the change in the bulk action is compensated by the change of the surface integral.

Here we study the behavior of a double trace fermionic operator in a thermal theory in the presence of a strong magnetic field, so first, in section \ref{sec:bulk}, we are concerned with the behavior of Dirac fields in the bulk, as we can be sure that a combination of them will be dual to the operator of our interest. It is important to mention that we will use a five dimensional background, and that nonetheless we can still use the correspondence because this background with its magnetic field is a solution to a consistent truncation of type IIB supergravity \cite{Uriel}. In section \ref{sec:flux} we carry an analogous calculation to that done in \cite{Laia:2011wf}, adapted to our background, starting by implementing all the ingredients necessary for the Wilsonian approach to renormalization, and proceeding then to determine the particular combination of fields that is dual to the operator we are interested in. Once we have the dual field, we apply to it the Wilsonian approach and find how the renormalization flux is affected by the reaction of the theory to the background magnetic field. We close that section describing the modifications that could arise by using a different renormalization scheme and identify the conclusions that can be draw independently of this choice. To provide further evidence for one of our conclusions, in section \ref{sec:DimRed} we make a very simple analysis of the behavior of the metric components to show that at low energy scales the plasma develops a substantially subluminal limiting velocity in the directions perpendicular to the background field, suggesting that the expected dimensional reduction takes place, and is maintained up to an energy scale that increases with the strength of the background magnetic field. We close with section \ref{sec:conclusions}, where we use our results to draw the conclusions stated in the abstract.

\section{The bulk equations}
\label{sec:bulk}
In this section we study a fermionic field in a five dimensional asymptotically AdS background with a constant magnetic field \cite{D'Hoker:2009mm} $F=Bdx\wedge dy$ turned on. The metric takes the form 
\begin{equation}
\label{eq:1.b}
ds_{_5}^2 = -U(r)dt^2 + V(r)\left( dx^2 + dy^2\right) + W(r) dz^2 + \frac{dr^2 }{U(r)},
\end{equation}
and we will pick the gauge $A_y =B x$ for the magnetic field. This background was introduced\-\cite{D'Hoker:2009mm} to model the gravitational dual of a gauge theory in the presence of a constant magnetic field in its $z$ direction, and has already been used to study a number of phenomena in the presence of such a magnetic field \cite{Arciniega:2013dqa,Arean:2016het,Ammon:2016szz,Ammon:2017ded}.

This background has to be constructed numerically, and the particular way in which we obtain it lends a family of solutions with a horizon located at $r=r_h$ and characterized by one single parameter $b$ that measures the intensity of the magnetic field. As the magnetic field vanishes, the elements of this family smoothly approach the black brane solution given by plugging
\begin{align}
\label{bb}
U_{BB}(r) &=(r+\frac{r_h}{2})^2(1-\frac{(\frac{3}{2}r_h)^4}{(r+\frac{r_h}{2})^4}),\nonumber\\
V_{BB}(r) &=\frac{4V_0}{9r_h^2}(r+\frac{r_h}{2})^2,\\
W_{BB}(r) &=\frac{4}{3}(r+\frac{r_h}{2})^2, \nonumber
\end{align}
in (\ref{eq:1.b}) and setting $b=0$.

For all elements of the family, except the one for $b=0$, the near horizon geometry is that of a BTZ black hole times a flat two dimensional space, given by inserting
\begin{equation} \label{btz}
U_{BTZ}(r)=3(r^2-r_h^2),\;\;\; V_{BTZ}(r)=\frac{B}{\sqrt{3}}\;\;\; \mathrm{and}\;\;\; W_{BTZ}(r)=3r^2,
\end{equation}
in (\ref{eq:1.b}), while the geometry close to the boundary is the asymptotic $AdS_5$ needed in the correspondence. As the intensity of the magnetic field increases, the transition from the near horizon geometry into the $AdS_5$ zone takes place at a larger radius. From this perspective we can think of the $b=0$ case as the member of this family of solutions for which the transition takes place right at the horizon.

We have already discussed elsewhere the specific way in which we carry this numerical construction, so we refer the interested reader to \cite{Arean:2016het}, and all we shall need to know about the background for the present work are the characteristics described in the previous paragraph and those that will be explicitly exhibit in section \ref{sec:DimRed}. In what follows, we will use the numerical results for $U,V$ and $W$ achieved from the explicit calculations.

We now consider the action
\be
S= \int_{r = r_h}^{r = \frac{1}{\epsilon}}d^5 x\sqrt{-g}\,{\cal L} + S_B[\psi,\bar{\psi},\epsilon],\label{action}
\ee
where the Lagrangian density for the fermionic field $\psi$ in the five dimensional background is given by
\begin{equation}
\label{eq:2}
{\cal L} = \bar{\psi}\left[ (1/2)\left( \Gamma^M \overrightarrow{D}_M -\overleftarrow{D}_M\Gamma^M\right) -m \right]\psi,
\end{equation}
and $S_B[\psi,\bar{\psi},\epsilon]$ is a boundary term that will be the topic of discussion bellow. For this work, $\psi$ will be considered to bare no charge to couple to the magnetic field so its minimal coupling to $A$ will be left out, and yet we will see that there is interesting physics in this first approach\footnote{The consistency of considering the fermionic field neutral with respect to $b$ can be seen in the uplift to ten dimensions of the five dimensional background we use here. The five dimensional effective field being neutral corresponds with the ten dimensional field being turned off in the internal directions, which can be consistently done \cite{Uriel}}. The grassmanian nature of the spinors in this action will be relevant for latter calculations. Also, throughout the paper, uppercase gammas are those that satisfy $\{ \Gamma^M,\Gamma^N\}=2g^{MN}$ for $g$ the metric associated to the line element (\ref{eq:1.b}), while lowercase gammas are those satisfying $\{\gamma^a,\gamma^b\}=2\eta^{ab}$, and hence they are related through $\Gamma^M = E^M_a \gamma^a$.
 
As in the expressions just above, we shall use an index notation where uppercase Latin letters refer to the coordinates of the spacetime with metric \eqref{eq:1.b}, $M= \{t,x,y,z,r\}$, whereas lowercase Greek indexes run only over the first four of these indexes $\mu =\{t,x,y,z\}$, leaving out the radial direction $r\in <r_h,\infty>.$ We reserve lowercase Latin indexes running from 0 to 4 for elements of the tangent or cotangent spaces that are written in terms of the tetrad
\begin{equation}
\label{eq:4}
\begin{split}
E^{t}_{a}  =\delta^0_a \,U^{-1/2} \,,
\qquad
E^{x}_{a} & =\delta^1_a \,V^{-1/2},\qquad  E^{y}_{a}  =\delta^2_a \,V^{-1/2}  \,,
\\
\\
E^{z}_{a}  =\delta^3_a \,W^{-1/2}, &\qquad  E^{r}_{a}  =\delta^4_a \,U^{1/2} \,,
\end{split}
\end{equation}
or its dual basis.

To compute the spin connection $\omega^{ab}_{M}  = E^a_P(\partial_M E^{bP} + E^{bQ} \Gamma^P_{MQ})$ needed for the covariant derivative operator $\overrightarrow{D}_M = \partial_M + \frac{1}{4}\omega^{ab}{}_M \left[\gamma_a, \gamma_b  \right]$ in (\ref{eq:2}), we first obtain all non-vanishing Christoffel symbols, which are given by 
\begin{equation}
\label{eq:3.a}
\begin{split}
\Gamma^t_{rt} & =\frac{1}{2} \frac{U^\prime}{U} \,,
\qquad
\Gamma^x_{rx}= \Gamma^y_{ry}= \frac{1}{2}\frac{V^\prime}{V} \,,
\qquad
\Gamma^z_{rz} =\frac{1}{2}\frac{W^\prime}{W} \,,
\\
\Gamma^r_{tt} & =+\frac{1}{2}{U} U^\prime \,,
\qquad
\Gamma^r_{xx}= \Gamma^r_{yy}= -\frac{1}{2}{U}V^\prime \,,
\qquad
\Gamma^r_{zz} = -\frac{1}{2}{U} W^\prime \, ,
\end{split}
\end{equation}
leading to the only non-vanishing components of the spin connection
\begin{equation}
\label{eq:5}
\omega^{bc}\mu =    U^b_\mu  \delta^c_4 -\delta^b_4 U^c_\mu , 
\end{equation}
where for compactness we have defined
\begin{equation}
\label{eq:6}
U^0_t =  \frac{1}{2} U^\prime, \,\, U^1_x = U^2_y = \frac{V^\prime}{2} \sqrt{\frac{U}{V}}  \,\,, U^3_z = \frac{W^\prime}{2} \sqrt{\frac{U}{W}},
\end{equation}
with all other $U^a_\mu=0$.

Even if the magnetic field makes the background anisotropic, there is still translational invariance in the four directions perpendicular to the radial one, so we can expand in plane waves and write
\begin{equation}
\label{eq:8}
\psi(x^\mu, r) = e^{i\omega t - i \vec{k}\cdot \vec{x}}\phi(r), \qquad k^\mu = (\omega,\vec{k}),
\end{equation}
so that the equation of motion that results from extremizing (\ref{action}) reads
\begin{equation}
\label{eq:9}
\left[ -i\gamma^\mu K_\mu(r) + U^{1/2}(r)\,\gamma^4\partial_4 - \frac{1}{2} U^{1/2}\gamma^4F(r) -m \right]\phi(r) = 0,
\end{equation}
with
\begin{subequations}
\label{10}
\begin{equation}
\label{10.a}
K^\mu \equiv \left(U^{-1/2}\omega, V^{-1/2}k_x, V^{-1/2}k_y, W^{-1/2}k_z\right),
\end{equation} 
and 
\begin{equation}
\label{10.b}
F(r) \equiv -\frac{1}{2}\left( \frac{U^\prime}{U} + 2\frac{V^\prime}{V} + \frac{W^\prime}{W}\right).
\end{equation} 
\end{subequations}

For the variation principle to be well defined, in the sense that a solution to the equation of motion (\ref{eq:9}) is guarantied to extremize (\ref{action}), we need to cancel the boundary contribution of any variation. We will do this here by imposing the Neumann boundary conditions
\be
\Pi=\frac{\delta S_B}{\delta \bar{\psi}}\,\,\,\,\, {\mathrm{and}} \,\,\,\, \bar{\Pi}=\frac{\delta S_B}{\delta \psi},\label{boundco0}
\ee
where $\Pi$ is the conjugated momentum of $\psi$ in the radial direction given by
\begin{equation}
\label{23}
\Pi \equiv -\frac{i}{2}\sqrt{-g} E^{r}_a\gamma^a\psi\,\, {\text{and}}\,\,
\bar{\Pi }  \equiv -\frac{i}{2}\sqrt{-g}\bar{\psi} E^{r}_a\gamma^a.
\end{equation}

The boundary term that respects covariance is given by
\begin{equation}
\label{bound0}
S_B = \frac{i}{2} \int_{r = \frac{1}{\epsilon}}d^4x \sqrt{-g_B}f\bar{\psi} \psi,
\end{equation}
where $g_B$ is the determinant of the metric induced at the surface ${r = \frac{1}{\epsilon}}$ that in our case is given by $V^2WU|_{r = \frac{1}{\epsilon}}$.

Noticing that $\sqrt{-g_B}=\sqrt{-g}E^{r}_4$ and that ${\gamma^4}^\dagger=\gamma^4$, both Neumann conditions (\ref{boundco0}) are reduced to ${\cal{M}}\psi=0$ and its conjugate for ${\cal{M}}\equiv(f+\gamma^4)$, where it is to be understood that an identity matrix is multiplying $f$. 

To determine the value that $f$ can take, we remember that a way to obtain information about an operator in the gauge theory using elements of the bulk physics, is to fix the value of the dual field at the boundary and extract the information we are looking for from the behavior of the fields conjugate momentum. Reversing the rolls in the previous paragraph of the value of the field and its momentum conjugated in the radial direction is also a possibility, and when more than one field, or field component, is involved, fixing some field values and some momentum values is operational just as long as half the boundary conditions are left unfixed.

The components of $\psi$ follow the first order equation (\ref{eq:9}), so the boundary conditions are fully determined by their values there, and hence, according to what was stated in the previous paragraph, only half the components of $\psi$ should be fixed at the boundary. Another way to say this is that the conjugate momentum to $\psi$ is also determined by its value.

If we want the condition (\ref{boundco0}), that we rewrote as ${\cal{M}}\psi=0$, to only fix the value of half the components of $\psi$, the matrix ${\cal{M}}\equiv(f+\gamma^4)$ has to project out half the components of $\psi$ which requires $f=\pm 1$. The previous statement can be seen to be true given that the square of $\gamma^4$ is the identity and hence the operators $P_+=\frac{1}{2}(1+\gamma^4)$ and $P_-=\frac{1}{2}(1-\gamma^4)$ satisfy the properties $i){P_\pm}^2=P_\pm$, $ii)P_\pm P_\mp=0$ and $iii)P_++P_-=1$, that is, they are a complete set of orthogonal projectors. The relationship $\gamma^a {P_+}\gamma^b=\gamma^a\gamma^bP_-$ also holds, and given that the kernel of any Dirac matrix is the empty set, they cannot change the rank of an operator by multiplying it, so the rank of $P_+$ and $P_-$ has to be the same and consequently, each of them projects out half the components of $\psi$, as anticipated.

We are left now with the decision of which one of the projected fields, $\psi_+\equiv P_+\psi$ or $\psi_-\equiv P_-\psi$, will be set to zero at the boundary, which is dictated by their asymptotic behavior as $r$ goes to infinity, that at leading order is the same as in the pure AdS case \cite{Laia:2011wf}, so here we only review the facts that will be useful for the calculations below.

In our background the AdS radius $L$ has already been fixed to 1, so in the limit $r\to \infty$ the three functions $U, V$ and $W$ approach $r^2$, while $F\to \frac{-4}{r}$ and $K_\mu\to\frac{k_\mu}{r}$. The region $r\to \infty$ is better explored using the coordinate $\rho=\frac{1}{r}$, in terms of which as we approach the boundary $\rho\to 0$, equation (\ref{eq:9}) becomes
\begin{equation}
\label{18}
\left[-i\rho\gamma^\mụ k_\mu+ \gamma^\rho \rho\partial_\rho - \frac{1}{2}\left( 4\gamma^\rho + 2m\right)\right]\phi(\rho) =0,
\end{equation} 
where $\gamma^\rho$ has to be equal to $-\gamma^r$ to keep $\Gamma^r\partial_r = \Gamma^\rho\partial_\rho$.

Whenever $m$ is not a half integer, the solution to this equation can be put in terms of the modified Bessel functions $I_\nu(k\rho)$ as 
\begin{equation}
\label{19}
\phi_{\pm}(k\rho) = \left( k\rho\right) ^{(\frac{5}{2})}\left[C^{\pm}_{\nu_{{}_\pm}}I_{\nu_{{}_\pm}}(k\rho) + C^{\pm}_{-\nu_{{}_\pm}}I_{-\nu_{{}_\pm}}(k\rho)\right],
\end{equation}
with $\nu_{\pm}= (m\mp1/2).$

For half integer $m$ these modified Bessel function are not linearly independent, so we take the Hankel function $K_\nu = (\pi/2 )i^{\nu + 1}H^{(1)}_\nu (ix)$ as a second solution, which asymptotically shows a characteristic logarithmic term.

Close to the boundary the fields are approached by
\begin{subequations}
\label{20}
\begin{equation}
\label{20.a}
\phi_-(k\rho)= A(k)\rho^{2 -m} + B(k)\rho^{2 + m + 1}  \cdots
\end{equation}
and
\begin{equation}
\label{20.b}
\phi_+(k\rho)= C(k)\rho^{2 +m} + D(k)\rho^{2 - m + 1}  \cdots
\end{equation}
\end{subequations}

The energy contribution of the asymptotic region evaluated using (\ref{20}) is given by the integral
\be
\int_{\rho=\epsilon}^{\rho=0}\frac{dr}{r^{4+1}}[\bar{A}Dr^{4-2m+1}-\bar{C}Br^{4+2m+1}],
\ee
so we see that for $m\geq 1/2$, the second term is normalizable while the first is not, hence for $A$ to be dynamical it would require an infinite amount of energy and this forces us to make $\psi_-=0$ keeping $\psi_+$ free. If $m\leq -1/2$ the situation is reversed, and the one that is left free is $\psi_-$. For $-1/2 < m < 1/2$ both terms are normalizable, so we can chose either field to be the one that stays dynamical.

The choices $\psi_-=0$ and $\psi_+ =0$ are respectively called standard and alternative quantization. These two options correspond to two fixed points in the renormalization flux on the gauge theory side and represent two different theories in which, as we shall see, if we set $\psi_-=0$, the operator in the gauge theory dual to $\psi_+$ has conformal dimension $\Delta_+ = 2 + m$ whereas for the alternative case the conformal dimension of the operator dual to $\psi_-$ has conformal dimension $\Delta_- = 2 - m$. We will keep the values of $m$ in the interval $-1/2 < m < 1/2$ and observe how the flow takes us from one to the other theory.

\section{The renormalization flux} 
\label{sec:flux}

We turn now to the renormalization group flow, and since we will still be working close to the boundary, we will keep using the coordinate $\rho$.

To begin with the Wilsonian approach to renormalization we need to start by integrating out the high energy degrees of freedom, in particular, those with energy in the interval $\{\Lambda+\delta\Lambda,\Lambda\}$, where $\Lambda$ sets the renormalization scale. The way to implement this on the gravity side this is to perform the integral
\begin{equation}
\label{21}
S[\epsilon +\delta\epsilon] -S[\epsilon] = \int_{\rho = \epsilon + \delta \epsilon}^{\rho = \epsilon}d^{d+1} x\sqrt{-g}\,{\cal L} + S_B[\psi(x,\epsilon + \delta \epsilon)] -S_B[\psi(\epsilon)].
\end{equation}

The Wilsonian approach to renormalization is translated to the gravity side \cite{Heemskerk:2010hk,Faulkner:2010jy} by observing that physical quantities should not depend on the position of the boundary, and consequently (\ref{21}) should not depend on $\epsilon$, condition which of course we will be only able to meet if we permit the boundary term to change as the boundary is moved. This flow of the boundary term with $\epsilon$ will encode the renormalization group flow as we will now see.

The boundary conditions (\ref{boundco0}) are to be imposed at $\rho=\epsilon$, since this is the radius associated to the renormalization scale and hence it should mark the boundary of our bulk. It would be convenient then if we write the variation of (\ref{21}) with respect to $\epsilon$ as an expression that is evaluated at $\rho=\epsilon$ solely, which we can do by using the fact that $\delta\epsilon$ is small and expand (\ref{21}) around $\rho=\epsilon$ to first order in $\delta\epsilon$ getting to
\begin{equation}
\label{22}
\frac{dS}{d\epsilon} = -\int_{\rho = \epsilon} d^{d} x\sqrt{-g}\,{\cal L} + \int_{\rho = \epsilon} d^{d} x\left(-\frac{\delta S_B}{\delta \psi}\partial_\rho \psi + \partial_\rho\bar{\psi}\frac{\delta S_B}{\delta \bar{\psi}}\right)
+ \frac{\partial S_B}{\partial \epsilon}
\end{equation}
where, as desired, all terms are evaluated at $\rho= \epsilon$.

Now that everything is evaluated at $\rho=\epsilon$, we can use the conditions (\ref{boundco0}) to write (\ref{22}) as
\begin{equation}
\label{24}
\frac{dS}{d\epsilon} = \int_{\rho = \epsilon}d^{d}x\,\,{\cal H} + \frac{\partial S_B}{\partial \epsilon},
\end{equation}
with the radial Hamiltonian given by
\begin{align}
\label{25}
{\cal H}  & = - \bar{\Pi} \partial_\rho \psi + \partial_\rho \bar{\psi}\Pi - \sqrt{-g}{\cal L}
\\
 & = -\frac{i}{2}\sqrt{-g} \left[E^{\mu}_a\bar{\psi}\left(2\gamma^a\partial_\mu + \frac{1}{4}\omega_{bc,\mu}\{ \gamma^a,[\gamma^b ,\gamma^c]\}  \right)  \right],
\end{align}
which is just the Legendre transformation of ${\cal{L}}$ in the radial direction.

For (\ref{21}) to be independent of the value of $\epsilon$ we need to see that its variation with respect to it, given by (\ref{25}), vanishes, condition that can be written as
\begin{equation}
\label{26}
\frac{\partial S_B}{\partial \epsilon} = - \int_{\rho = \epsilon}d^d {\cal H},
\end{equation}
which is the Hamilton-Jacobi equation dictating the flow of the boundary term, dual to the Callan-Zymanzyk equation.

For our metric, as any other diagonal metric with components depending only on the radial direction, $E^\mu_a\omega_{bc,\mu}\{ \gamma^a,[\gamma^b ,\gamma^c]\} =0$, and, given that $\partial_\mu \psi=0$, the flow equation can be written as
\begin{equation}
\label{28}
\frac{\partial S_B}{\partial r} = i \frac{L^2}{r^2}\int_{r= 1/\epsilon}d^d x\sqrt{-g}\left[m \,\bar{\psi}\psi \right],
\end{equation}
where we have returned to the coordinate $r=1/\rho$.

It will be relevant to close this section by noticing that all we did in it is independent of the particular form of the boundary action, and so anything stated here will apply to the deformed theory that we will study in the following section.

\subsection{Deforming the theory}

As originally stated, one of the things we are interested on is the exploration of the impact that turning on an intense magnetic field on a theory at finite temperature would have on the fermionic renormalization flow. As a trial case, we will use the deformation of the theory given by the relevant operator
\be
\Delta S_{\mathrm{Dirac}}=i\int\frac{d^3k}{(2\pi)^3}\xi\bar{\Psi}(k)\Psi(k), \label{diracdef}
\ee
with a constant $\xi$ studied in \cite{Laia:2011wf} so that we can compare our results with the ones there, and in  particular, recover them when we set $b=0$.

For defines, we will use the representation of the Dirac matrices given by
\small
\begin{equation}
\label{eq:matrix}
\gamma^0 = \left(\begin{array}{cc}
0&1\\
-1&0
\end{array}  \right),
\gamma^1 = \left(\begin{array}{cc}
0&\sigma^1\\
\sigma^1&0
\end{array}  \right),
\gamma^2 = \left(\begin{array}{cc}
0&\sigma^2\\
\sigma^2&0
\end{array}  \right),
\gamma^3 = \left(\begin{array}{cc}
0&\sigma^3\\
\sigma^3&0
\end{array}  \right),
\gamma^4 = \left(\begin{array}{cc}
1&0\\
0&-1
\end{array}  \right),
\end{equation}
\normalsize
and also remember that in five dimensions there is no $\gamma^5$ matrix, but that nonetheless, in the four dimensional space of the gauge theory, this roll will be assumed by $\gamma^4$, that is, $\gamma^5_{(4-dim)}=\gamma^4$.

Even before determining the appropriated boundary term, we already see that after the conditions (\ref{boundco0}) have been imposed, we need an extra fermionic field in the bulk, say $\chi$, so that along with $\psi$ they provide enough degrees of freedom to encode those of the four components of the fermionic operator $\Psi_\mu$ in the gauge theory. Just like $\psi$, $\chi$ will also have an expansion identical to (\ref{20}), except with its own fermionic operators, that we will call $\tilde{A},\tilde{B},\tilde{C}$ and $\tilde{D}$ just to tell them apart. As we did with $\psi$, we want to impose on $\chi$ the boundary conditions that correspond to the alternative quantization, so the necessity for the conformal dimension to be equal for all the components of $\Psi_\mu$ demands for $\chi$ to have the same mass as $\psi$.

To model (\ref{diracdef}) we need to introduce, along with (\ref{bound0}) and the corresponding expression for $\chi$, a boundary term using $\psi$ and $\chi$ in a way in which the result has the right symmetries and properties. The total term turns out to be
\begin{equation}
\label{29}
S_B = \frac{i}{2} \int_{\rho = \epsilon}d^dx \sqrt{-g_B}\left[ f\left(\bar{\psi} \psi+ \bar{\chi}\chi\right) + g\left(\bar{\psi}^{(c)}\chi+ \bar{\chi}\psi^{(c)}\right)\right],
\end{equation}
where $\psi^{(c)} \equiv \gamma^2\psi$ is the charge conjugate of the spinor $\psi$. Notice that we have multiplied the first two terms by the same constant $f$, and we did this since different components of $\Psi$ will come from $\psi$ and $\chi$, so in this way $\Psi$ will transform correctly under the Lorentz group.

About this boundary term we notice that since under a chiral symmetry transformation $\psi \to e^{i\alpha}\psi$ and $\bar{\psi}^{(c)}\to e^{-i\alpha}\bar{\psi}^{(c)}$, the second term in (\ref{29}) breaks chirality. Also, in odd spacetime dimensions, the pin group $Pin(1,d)$ is associated with the {\it twisted map}, that sends odd elements of the Clifford algebra to minus themselves, and so for our five dimensional case, expressions like a mass term of the type $\bar{\psi}\psi$ contained in (\ref{29}) break parity. Let us remember that the leading order in both, (\ref{20}) and the corresponding expansion for $\chi$, are those proportional to the expected value of the operator $\Psi$, so the final thing to notice is that since the terms multiplying $g$ in (\ref{29}) are proportional to $A^\dagger\gamma^0\tilde{A}$ and $\tilde{A}^\dagger\gamma^0A$, they are dual to an operator with two copies of $\Psi$, which is the nature of the double trace operator that we are looking for\footnote{This proportionality can be seen by noticing that for our choice of gamma matrices, $\gamma^0\gamma^2\sim \text{diag} (\sigma_2, \sigma_2)$.}.

The considerations just made, make it so that the boundary term we added is dual to a double trace operator that breaks chirality, making it a likely candidate to model (\ref{diracdef}). 
 
Given (\ref{29}), the boundary conditions (\ref{boundco0}) now read
\begin{subequations}
\label{30}
\begin{equation}
\label{30.a}
\left(f +\gamma^4\right) \psi = -g\gamma^2\chi,
\end{equation}
and 
\begin{equation}
\label{30.b}
\left(f +\gamma^4\right) \chi = -g\gamma^2\psi.
\end{equation}
\end{subequations}
By applying $\left(f -\gamma^4\right)$ on (\ref{30.a}) and using (\ref{30.b}) or the other way around, we see that the condition
\begin{equation}
\label{31}
f^2 + g^2 = 1,
\end{equation}
has to be satisfied.
 
This result is independent of the metric, and it will be so as long as the metric is diagonal and depends only on the radial coordinate, which are conditions satisfied in particular by pure AdS.

\subsection{The RG flow.}

As argued in the previous section, the term multiplying $g$ in (\ref{29}) is dual to the double trace operator in (\ref{diracdef}), so the renormalization flow of the coupling constant of the latter will be encoded in the Hamilton-Jacobi equation for $g$,
\begin{equation}
\label{33}
\left[ \partial_r \left( V\sqrt{WU}f\right) \left( \bar{\psi}\psi + \bar{\chi}\chi\right) + \partial_r \left( V \sqrt{WU} g\right) \left(\bar{\psi} \gamma^2\chi + \bar{\chi}\gamma^2 \psi\right) \right] = 2m \left( \bar{\psi} \psi + \bar{\chi}\chi\right),
\end{equation}
obtained by substituting the expression for the boundary action (\ref{29}) into (\ref{28}).

To polish of (\ref{33}), we see that manipulating the boundary conditions (\ref{30}) we also get
\begin{subequations}
\label{32}
\begin{equation}
\label{32.a} 
f\left( \bar{\psi}\psi + \bar{\chi}\chi\right) + g\left( \bar{\chi}\gamma^2\chi + \bar{\psi}\gamma^2 \chi\right) =0,
\end{equation}  
and
\begin{equation}
\label{32.b}
\bar{\psi}\left( f -\gamma^4\right) \psi + g\bar{\chi} \gamma^2\psi =0  \qquad \bar{\chi}\left( f -\gamma^4\right) \chi + g\bar{\psi} \gamma^2\chi =0,
\end{equation}
\end{subequations} 
that can be used in (\ref{33}) to either eliminate the bilinear $\left( \bar{\psi}\psi + \bar{\chi}\chi\right)$ in favor of $\left( \bar{\chi}\gamma^2\chi + \bar{\psi}\gamma^2 \chi\right)$, or the other way around, to then use (\ref{31}) and be left with
\begin{equation}
\label{34}
-\sqrt{U}\left( \partial_r g\right) = 2mg( \pm\sqrt{1-g^2}).
\end{equation}

We will see that if we start at $r=0$ with the negative sign for the square root, the flow takes $g$ from zero to 1 at some finite value of $r$, and after this point it is necessary to take the square root with the positive sign to carry with the flow that now takes $g$ back to zero as $r$ approaches infinity.

This flow is easier to follow in the equation for $f$,
\be
\sqrt{U}\left( \partial_r f\right) = 2m \left({1-f^2}\right)\label{fflow}
\ee
that, given the restriction $f^2+g^2=1$, is totally equivalent to (\ref{34}), and does not have a square root, so we do not need to pick the sign for different regions and smoothly takes $f$ from -1 for $r=0$ to +1 for $r\to\infty$. 

Given that all integrations in the following section will be performed numerically, in practice we will use the flow equation for $f$ to do the calculations.

\subsection{The effect of the magnetic field}

As anticipated in the introduction, so far we have determined which results obtained in pure AdS carry to our background, and now we are ready to explicitly see the impact that the magnetic field has in the gauge theory.

To provide a reference we notice that in AdS, equation (\ref{34}) reeds $-r\left( \partial_r g\right) = 2mg \sqrt{1-g^2}$, with solution
\be
g_0=\frac{4\xi r^{-2m}}{4+\xi^2r^{-4m}},\label{gsolads}
\ee
where following \cite{Laia:2011wf}, we have written the solution so that the coupling constant $\xi$ in (\ref{diracdef}) appears as an integration constant.

In general, given that as we already mentioned, the dimension of $\Psi$ is $\Delta_-=2-m$, our double trace operator has dimension $4-2m$, and its coupling constant in the ultraviolet limit will be given by the value of $g(r)r^{2m}$ as $r$ goes to infinity.

To follow the flow now, we can either fix the physics in the infrared, equivalent to fixing the value of $g$ for small $r$, and observe the flow as we move towards the ultraviolet for different values of the intensity of the magnetic field $b$, or fix the physics in the ultraviolet, equivalent to fixing the value of $g$ for large $r$, and see how the flow goes as we lower the energy, again, for different values of $b$. One subtlety is that, since we are not working on pure AdS, the energy scale is not directly given by $r$ but, for the particular form of our metric, it should be given by $\mu=U^{1/2}$. Consequently, working at a fixed energy scale, means working at slightly different radios.

\begin{figure}
\includegraphics[scale=.4]{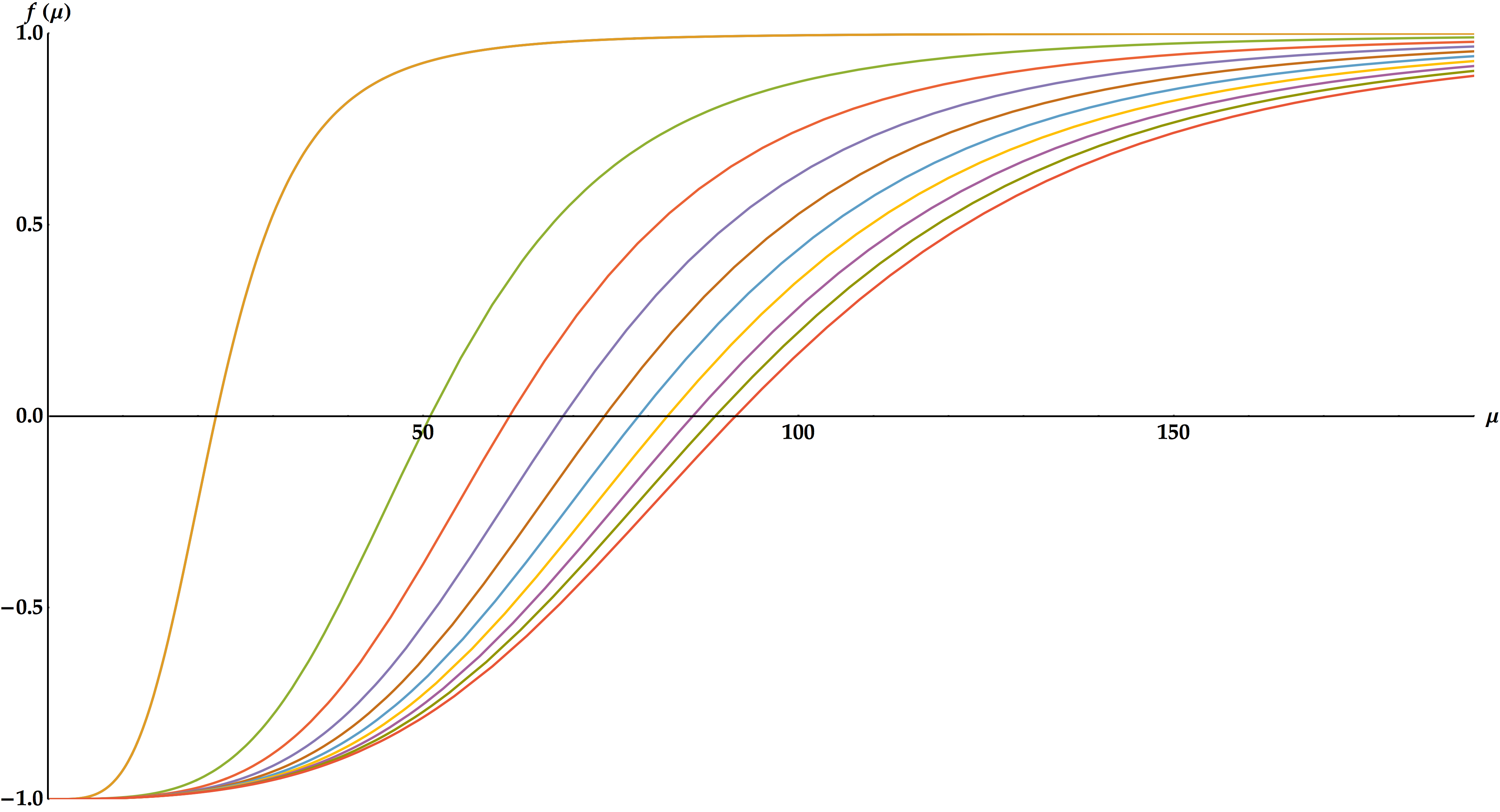}
\caption{$f$ as function of the energy scale $\mu$ for $b=\{0, 679, 1681, 2863, 4180, 5608, 7130, 8735, 10415, 12164\}$, where increasing values of $b$ are further to the right with the first plot on the left given by the analytic result for the pure AdS case. All the plots share the same value for $f$ at a particular infrared energy scale.}
\label{fir}
\end{figure}

\begin{figure}
\includegraphics[scale=.4]{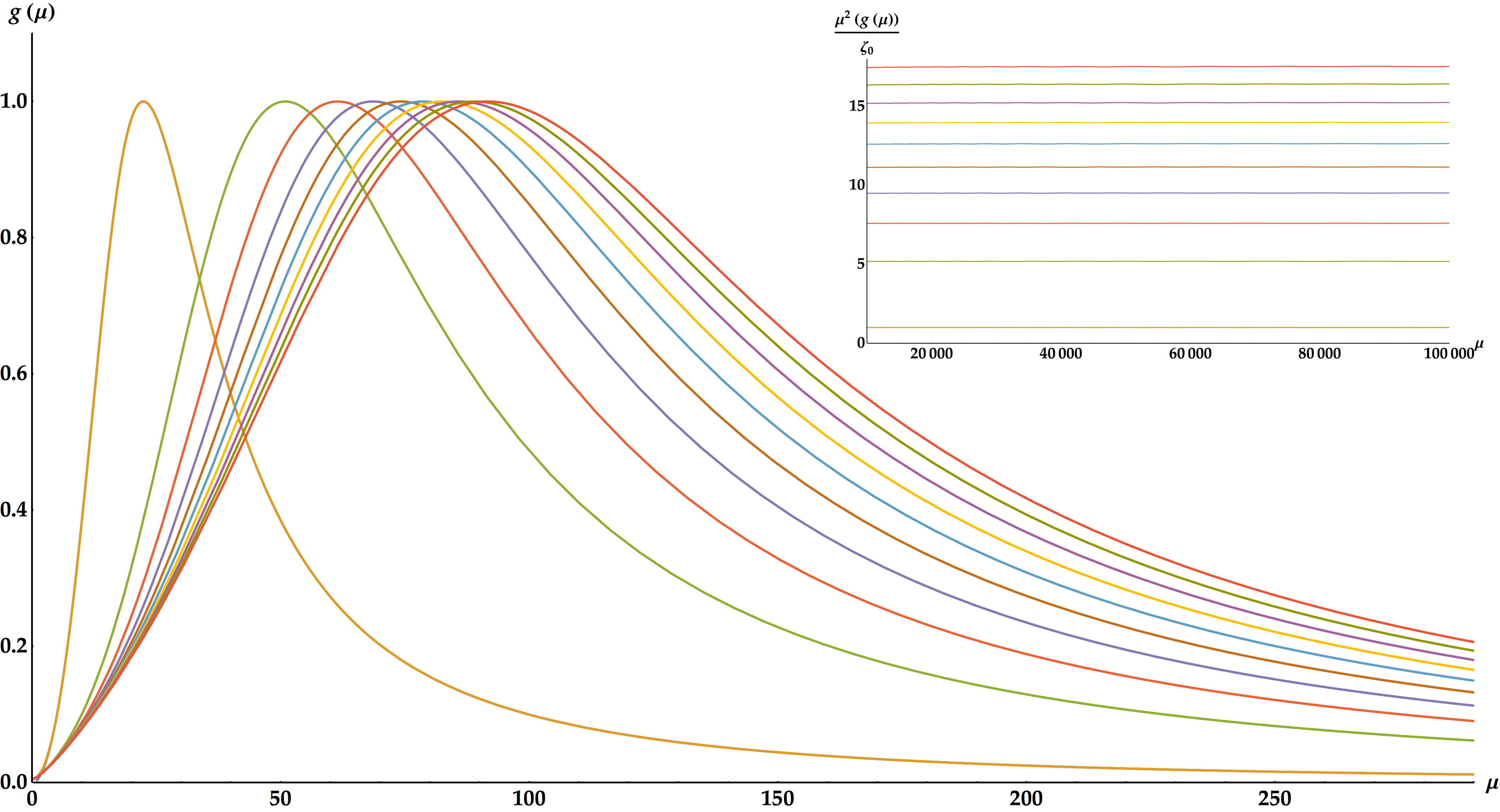}
\caption{g as function of the energy scale $\mu$ for $b=\{0, 679, 1681, 2863, 4180, 5608, 7130, 8735, 10415, 12164\}$, where increasing values of $b$ are further to the right with the first plot on the left given by the analytic result for the pure AdS case. In the inset we show how for each value of $b$, $\mu^2 g$ approaches a different constant that corresponds to the coupling of the gauge theory in the ultraviolet fixed point, that has been normalized with respect to the value $\xi_0$ in pure AdS. All the plots share the same value for $g$ at a particular infrared energy scale.}
\label{gir}
\end{figure}

\begin{figure}
\includegraphics[scale=.4]{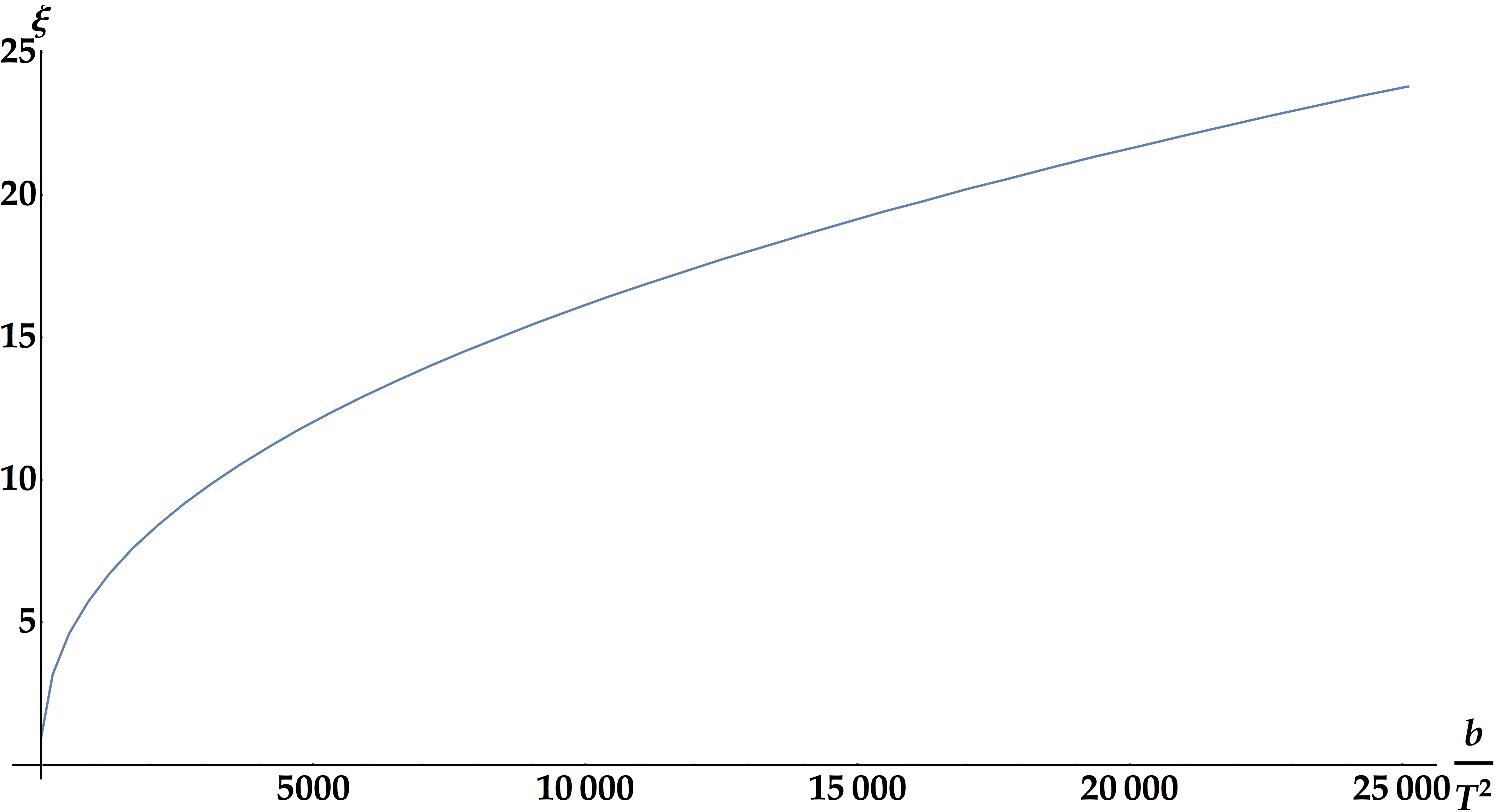}
\caption{The coupling constant of the gauge theory in the ultraviolet fixed point as a function of the intensity of the magnetic field with the value of $g$ fixed to a constant at a particular infrared energy scale for all values of $b$.}
\label{coup}
\end{figure}

We start with the first of these alternatives and depict the results in figures (\ref{fir}) to (\ref{coup}).

In figure (\ref{fir}) and (\ref{gir}) we respectively depict $f$ and $g$ as functions of the energy scale for the values of $b=\{0, 679, 1681, 2863, 4180, 5608, 7130, 8735, 10415, 12164\}$. The first line to the left of both plots is the $b=0$ analytic result for either $g_0$ given by (\ref{gsolads}), or $f$ given by
\be
f_0=\frac{4-\xi^2r^{-4m}}{4+\xi^2r^{-4m}}.\label{fsolads}
\ee

We will just make the obvious observation here that if we fix the physics in the infrared, the presence of the magnetic field makes the transition to the other fixed point to happen at a higher energy scale, and leave a more extensive discussion for latter. In (\ref{coup}) we plot, as a function of $b$, the value that the coupling constant $\xi$ takes at the ultraviolet fixed point, obtained as $\xi=$lim$_{r\to\infty} g(r)r^{2m}$ for the flow with the corresponding value of $b$. This function is monotonic in $b$, but far from linear.

\begin{figure}
\includegraphics[scale=.4]{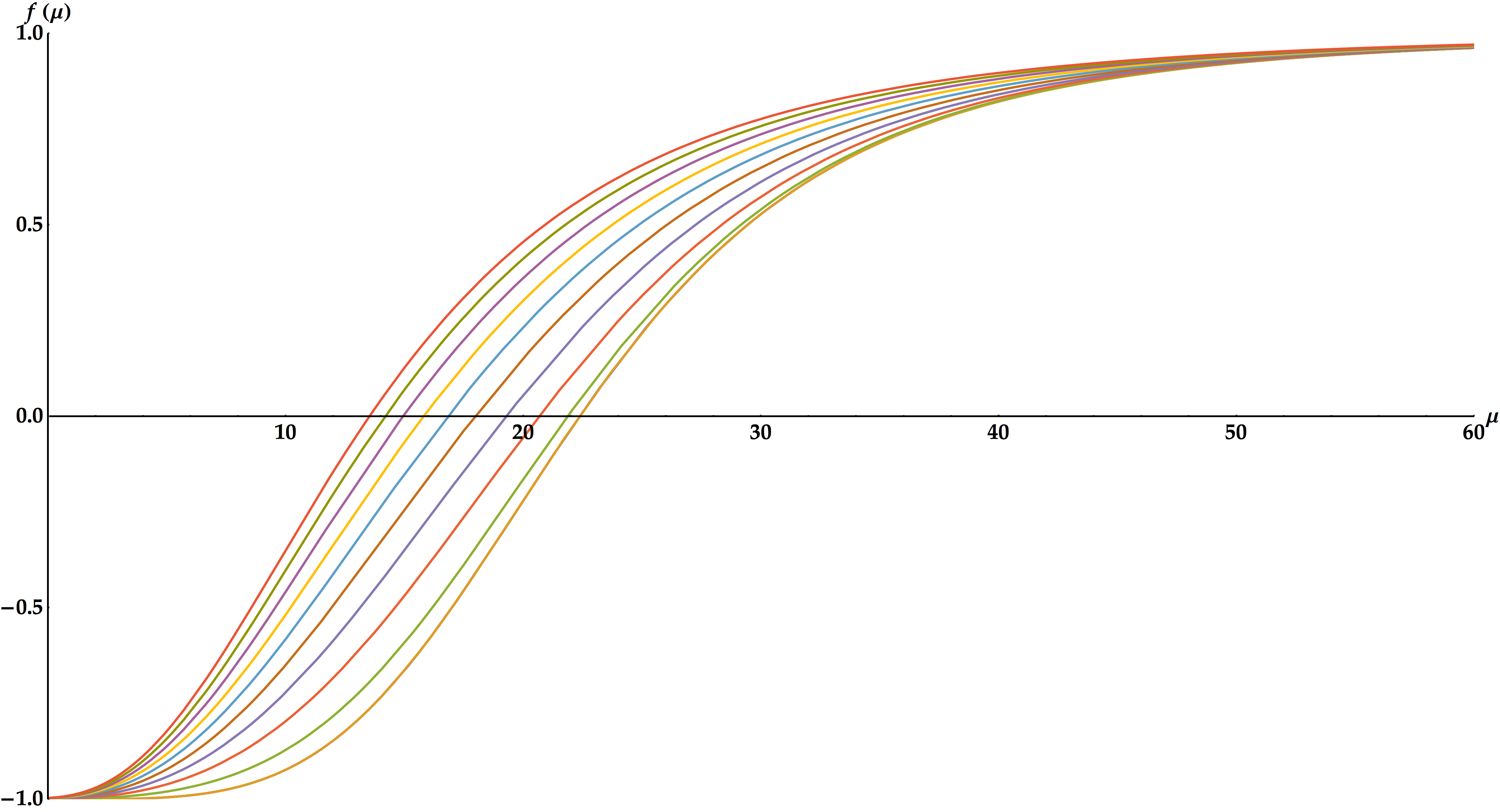}
\caption{$f$ as function of the energy scale $\mu$ for $b=\{0, 679, 1681, 2863, 4180, 5608, 7130, 8735, 10415, 12164\}$, where increasing values of $b$ are further to the left with the first plot on the right given by the analytic result for the pure AdS case. All the plots share the same value for $f$ at a particular ultraviolet energy scale.}
\label{fuv}
\end{figure}

\begin{figure}
\includegraphics[scale=.4]{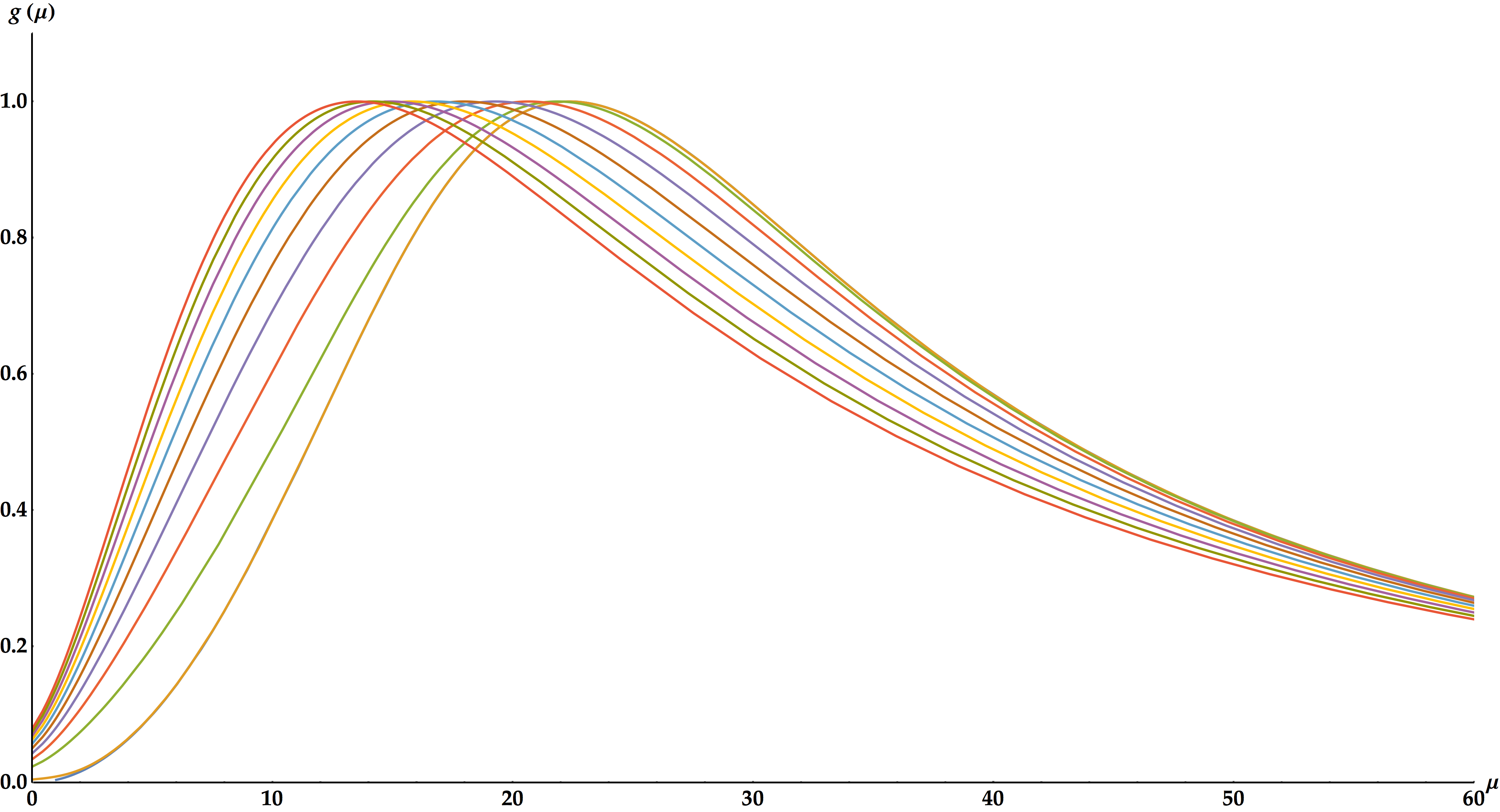}
\caption{g as function of the energy scale $\mu$ for $b=\{0, 679, 1681, 2863, 4180, 5608, 7130, 8735, 10415, 12164\}$, where increasing values of $b$ are further to the left with the first plot on the right given by the analytic result for the pure AdS case. All the plots share the same value for $g$ at a particular ultraviolet energy scale.}
\label{guv}
\end{figure}

The result of the second alternative, that of fixing physics in the ultraviolet, is depicted in figures (\ref{fuv}) and (\ref{guv}) that are analogous to (\ref{fir}) and (\ref{gir}) with the same values for $b$. In this case we do not show a plot like (\ref{coup}) because the value of $\xi$ has precisely been kept the same for all values of $b$ in the ultraviolet fixed point.

This time the plots of the analytic results for $b=0$ lie to the right of the plots for higher values of $b$. Consistent with our observation about the transition to the ultraviolet fixed point happening at higher energy scales for higher values of $b$ when things are fixed in the infrared fixed point, here we notice that when we keep things fixed in the ultraviolet point, the transition to the infrared theory happens at lower energy scales for higher values of $b$.

\subsection{Dependence on the renormalization scheme}
\label{Scheme}

The calculations in the previous subsection have been carried out in a particular renormalization scheme, which in \cite{Balasubramanian:2012hb} is refereed to as "maximal", since all contact terms have been removed. To draw any meaningful conclusions we need to determine the effect that a change of scheme could have in our results.

A change of scheme is a redefinition of the operators and the coupling constants of a theory at short distances, and consistently, in \cite{Balasubramanian:2012hb} it was shown that the difference between the scheme we use and others, like the minimal scheme, is given by the addition of contact terms, that is, $\delta-$function singularities at coincident points. In \cite{Balasubramanian:2012hb} it was also shown that the inclusion of these contact terms in momentum space is given by the addition of analytic functions on the boundary term. For the fermionic theory we are working with, these additions, and some more general ones, were considered in \cite{Laia:2011wf}, reaching the conclusion that they could be reabsorbed by allowing $g$ and $f$ to depend on the $k^\mu$ in (\ref{eq:8}).

Now that we know what a change of scheme amounts to, let us determine the repercussions of performing one.

We see that the behavior of $f$ and $g$ with respect to $r$ is unchanged, since even if $f$ and $g$ are invested with a dependence on $k^\mu$, equations (\ref{34}) and (\ref{fflow}) remain the same, as they are differential equations on $r$.

As noticed in \cite{Balasubramanian:2012hb}, another consequence of adding these contact terms is that the Dirichlet boundary conditions that in the previous subsection were satisfied at the radius dual to the energy scale, would have to be modified to be mixed, Newman and Dirichlet, boundary conditions to recover the particular solutions we studied here. This implies that the assumption that $g$ provides the coupling constant of the double trace operator while its expectation value is read from the derivative of $g$ is not necessarily valid and some mixing in this relationship is expected. As pointed out in \cite{Balasubramanian:2012hb}, for this mixed boundary conditions energy is not conserved, which can be explained by the mixing of single and double trace operators. All of this put together is to say that the coupling constant $\xi$ of the double trace operators for an energy scales/radius in between fixed points is a general function of $g, f$ and their derivatives with respect to $r$, $\xi(f,f',g,g')$, and that the explicit dependence of this function on its arguments is fixed by the renormalization scheme. The behavior depicted in figures (\ref{fir}, \ref{gir}, \ref{fuv}, \ref{guv}) indicates that both $g,$ and $f$ have two regions of well defined characteristics, one for the infrared and another one for the ultraviolet, which correspond to the fixed points of the theory. Furthermore, in the same figures we see that as the intensity of the magnetic field is increased, the separation in the energy scale between the two asymptotic regions grows. Now, if we were to study the crossover from one fix point to another by using the actual coupling constant $\xi(f,f',g,g')$ in a general scheme, we would still observe that the separation in the energy scale between the asymptotic regions would increase, since it does for all the arguments of $\xi$, leaving it with no option but to follow this behavior since its dependence on them is fixed by the scheme. The previous conclusion is in agreement with the expectation from the field theory side that, since the space in which the renormalization flow occurs has only one parameter that can modify it, which is the intensity of the magnetic field, its topology should not change, since for this to happen, there should be a scheme in which the lines that represent a particular flow, should either cross, which they cannot because they are integral lines of a flow, or should fully overlap, changing the dimension of the renormalization space, which again, cannot happen\footnote{We would like to thank David Berenstein for pointing this out.}.

It is true though, that there could be a scheme in which the shape of the plots for $\xi$ differs so dramatically from figures (\ref{fir}, \ref{gir}, \ref{fuv}, \ref{guv}) that it could hide the pattern just described, or diminish it enough so that is not evident, but we would expect this to be rare. Of course the only way to fully corroborate what we have just stated is to compute a number of scheme independent quantities that would make the delay of the crossover irrefutably universal, but we will leave this calculation for the future and will provide just one example in section \ref{sec:DimRed}.

From our current results, the one that is scheme independent from the start is figure (\ref{coup}), since this coupling constant $\xi$ is obtained from the behavior close to the ultraviolet fixed point, where no ambiguity exists between the coupling and the expected value, nor between single and double trace operators, as the constant we are looking for comes from the mode in $g$ and $f$ that goes as $r^{-2m}$ as $r\rightarrow\infty$.
 
\section{Subluminal limiting velocity and the dimensional reduction of the gauge theory}
\label{sec:DimRed}

Even though the arguments stated in subsection \ref{Scheme} are robust enough to ensure that the behavior that we have reported so far is going to be qualitatively present in all renormalization schemes, it would of course be more satisfactory to provide an scheme independent calculation that shows that the crossover from the infrared theory to the ultraviolet one actually happens at higher energy scales as the intensity of the background field is increase. 

One such calculation is to compute the amplitudes for fermion scatterings in the background we have used so far, determining first a clear distinction between low and high energy amplitudes, and then, verify that the spread in the characteristic energies for one or the other type of amplitude to occur gets larger with the intensity of the background field. This calculation, along with other dynamical ones, would be the object of future work, but for the time being, a scheme independent phenomenon we want to study to provide further evidence of our results is the dimensional reduction that a theory like the one we are working with should experience at low energies, so we proceed to explore this through the its causal structure.

The causal structure in a gauge theory is given by the conditions that have to be satisfied by the points at which two fields are evaluated so that these fields either commute for the bosonic case or anticommute for the fermionic one. The commutators or anticommutators are normally written in terms of the subtraction or addition of propagators,  and hence contain relevant information of the theory. Even if the propagators themselves depend in the renormalization scheme, the causal structure does not, so determining this structure provides scheme independent information. A way to extract the causal structure in our context is to use an arbitrary renormalization scheme to compute the propagation amplitudes for the fermion fields, perform the necessary addition and find the conditions for the result to vanish. This is not the way we will proceed here, since it is much simper to determine the shape of the lightcones at each energy scale by direct inspection of the speed of light in different directions, and as we will explain below, it is equivalent.

We know in advanced that we will find a qualitatively different causal structure in the infrared and the ultraviolet limits of the theory we are working with, since the dimensional reduction that a gauge theory should undergo in the presence of a strong magnetic field has already been studied using the gauge/gravity correspondence. For instance in \cite{D'Hoker:2016wgl} a formal analysis was performed to exhibit how the operator algebra is projected to the one that a lower dimensional theory should have.

More recently, in \cite{Arean:2016het}, we performed a dynamical calculation to show that the drag force that a particle experiences when traveling in directions perpendicular to the magnetic field, increases linearly and without a bound as the intensity of the field grows bigger, while it stays bounded for propagation along the field. We obtained part of the results in \cite{Arean:2016het} by studding the motion of a string embedded in the same background used here \ref{eq:1.b}, and determining when the world sheet develops a horizon as a consequence of the dependence on the radius of the local speed of light. The results in \cite{Arean:2016het} indicate that the causal structure obtained by this dynamical calculation is determined by the shape of the lightcones at a specific radial location, making it a structure that depends on the energy scale and that can be read directly from the metric. In \cite{Arean:2016het} we did not investigate the dependence of the results in the energy scale, since it was not the objective at the time.

In what follows we will analyze how the presence of a magnetic field causes the plasma to develop a substantially subluminal limiting velocity in the directions perpendicular to it, leading to another indication of the dimensional reduction just discussed, but more importantly, we will show that the energy scale up to which this reduction is effective increases with the intensity of the background magnetic field. We need to point out that the dimensional reduction is proper of the field theory itself, as it is experienced by the plasma that provides the vacuum of our theory, made by fields of which some are charged with respect to the magnetic field, and hence, they are subject to the physics alluded in \cite{D'Hoker:2016wgl} and \cite{Arean:2016het}. The fermions we added in the present work are not charged with respect to the magnetic field, so they will be affected by the dimensional reduction through the interaction with the vacuum of the theory that is reacting to the presence of the magnetic field, and not by direct coupling.

In our case, it is important to notice that the constant magnetic field makes the dual gauge theory anisotropic by singling out the direction in which it points, and therefore the propagation of particles does not need to be isotropic. From the very simple form of the metric we are using, we see that the proper speed of light in a given direction $x_i$ at radius $r$ is given by $\tilde{c}_i(r)=\sqrt{\frac{-g_{00}(r)}{g_{x_ix_i}(r)}}$. Locally, at any point of the bulk, this effect is of course unperceivable as all observations are due exclusively to the particulars of local coordinates. Nonetheless the holographic projection makes it so that when working in the gauge theory, where coordinate velocities are used, at an energy scale $\mu$ corresponding to a certain radius $r^*$, $\tilde{c}_i(r^*)$ dictates a limiting velocity. The way in which a limiting velocity appears has already been discussed in \cite{Argyres:2006vs}, while an argument  about its validity along with its computation for the case of a single quark can be found in \cite{Argyres:2008eg,Chernicoff:2008sa,Chernicoff:2011xv} and a microscopic description is done in \cite{Mateos:2007vn,Ejaz:2007hg}.

In the following few paragraphs we will show that when working at low energy scales the limiting velocity in directions perpendicular to the magnetic field are very small in comparison to the one along it, which even at those low energies approaches the speed of light.

\begin{figure}[!htb]
\minipage{.99\textwidth}
  \includegraphics[width=\linewidth]{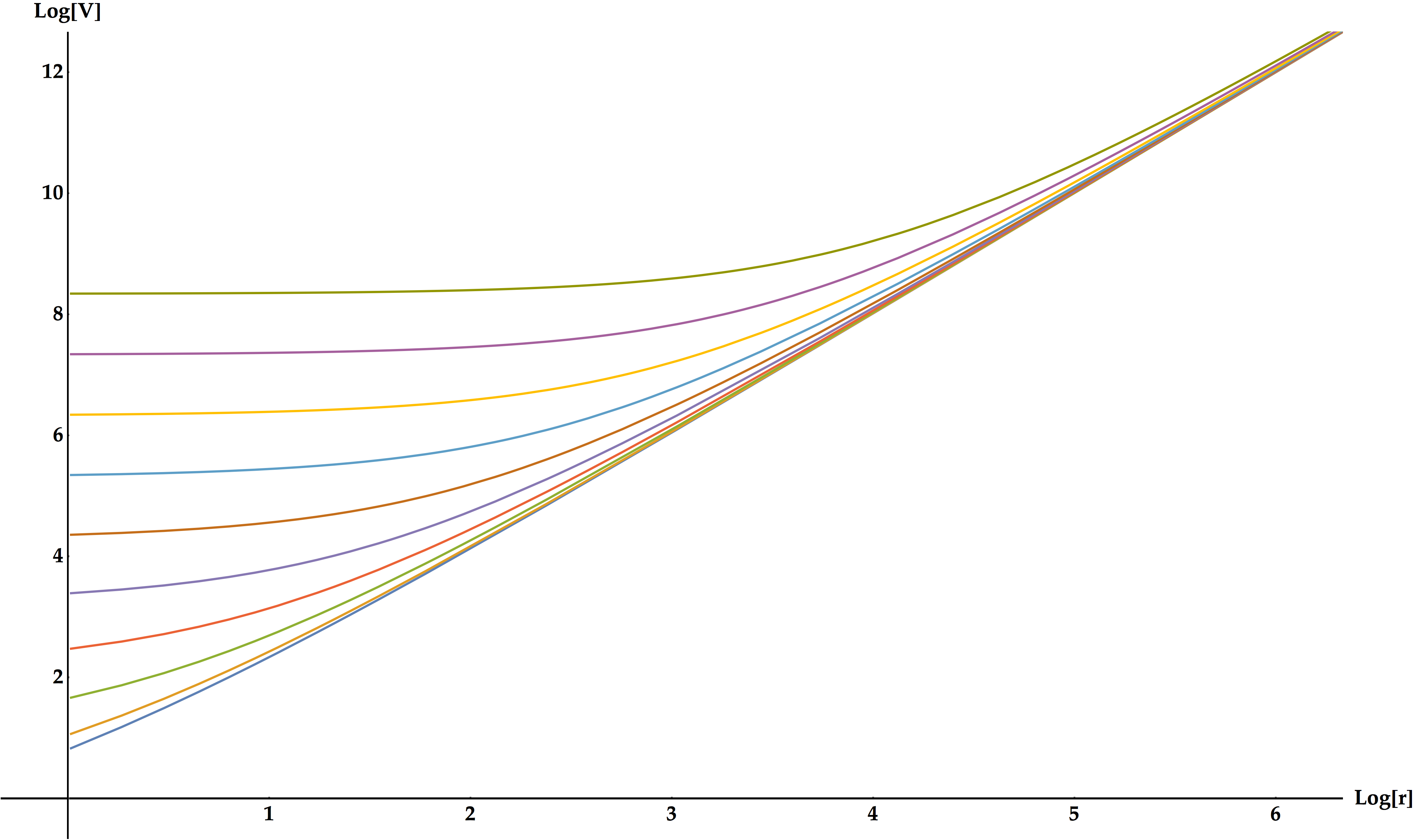}
\endminipage\hfill
\minipage{0.49\textwidth}
  \includegraphics[width=\linewidth]{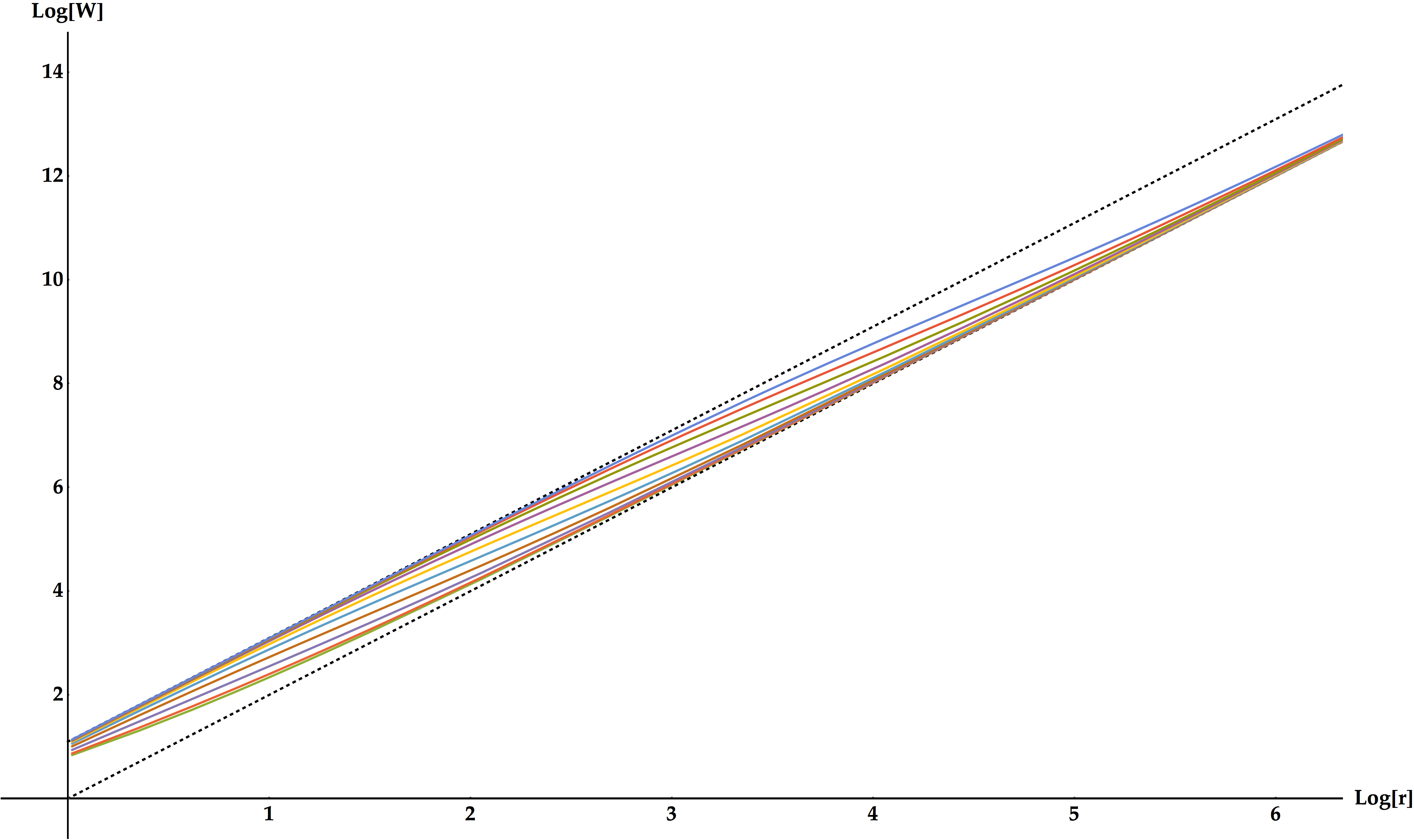}
\endminipage\hfill
\minipage{0.49\textwidth}%
  \includegraphics[width=\linewidth]{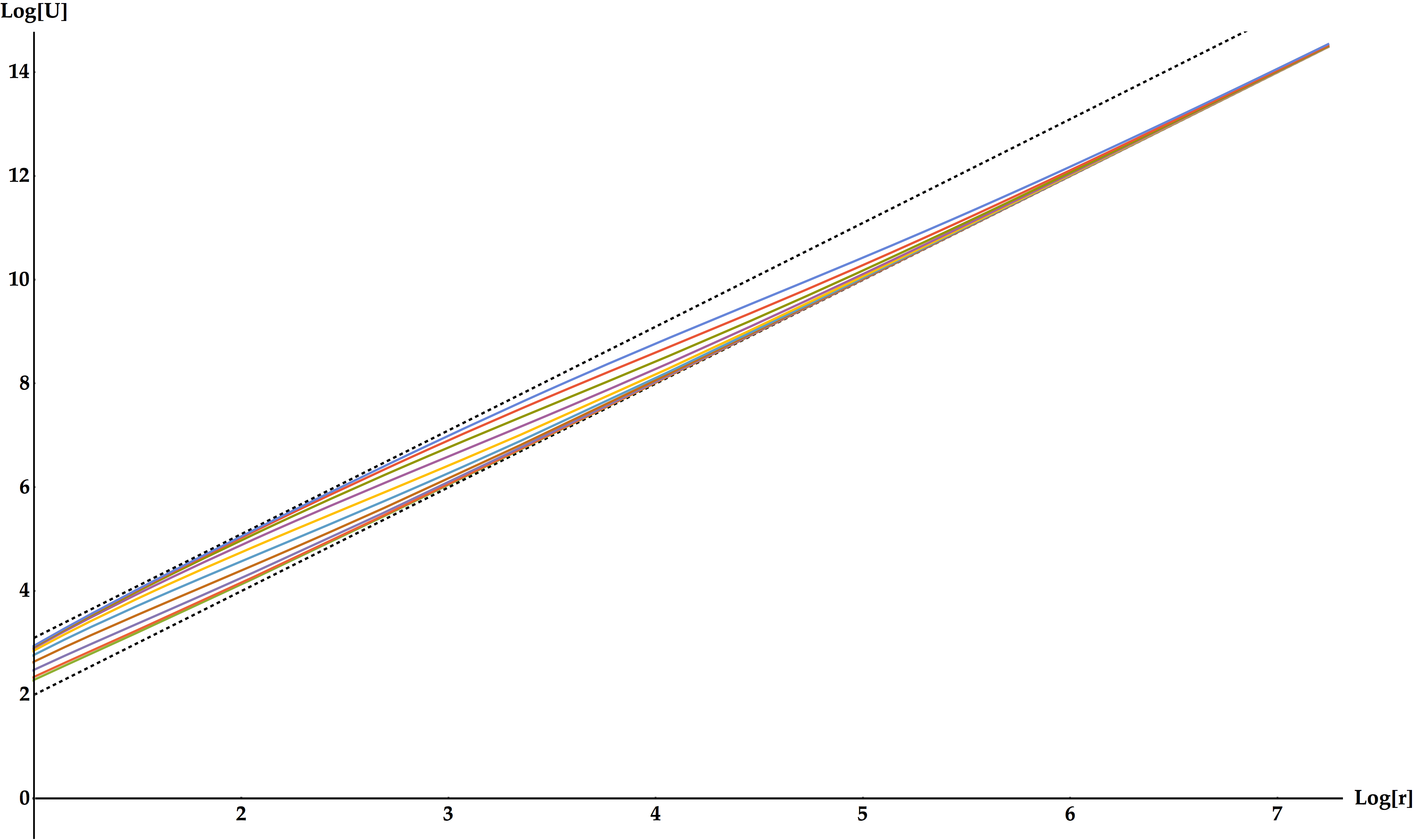}
\endminipage
\caption{Logarithm of the metric functions vs. logarithm of $r$ for $b=\{ 0, 8.62,
26.81, 75.05, 207, 570, 1564, 4277, 11680, 31860, 86861\}$. $V$ in the first plot shows how it starts as a large constant close to the horizon and it transitions into going like $r^2$. $W$ in the second plot shows how it starts as $3r^2$, shown as one of the dotted lines, close to the horizon and it transitions into going like $r^2$, shown as the other dotted line. $U$ in the third plot shows how it starts in zero, behaves like $3r^2$, shown as one of the dotted lines, for some intermediate values of $r$ and then transitions into going like $r^2$, shown as the other dotted line. The radius at which the transition happens for the three metric coefficients increases with the intensity of $b$.}\label{logback}
\end{figure}

To begin the analysis let us remember that the background we are working with transitions from the near horizon geometry (\ref{btz}), in which $V$ is a constant while $U$ and $W$ grow as $r^2$, to AdS$_5$, where, for the coordinates we are using, all the metric functions go like $r^2$. As can be seen in the logarithmic plots (\ref{logback}) the radius, and hence the dual energy scale, at which this transition occurs grows larger as the intensity of the magnetic field is increased. It is therefore possible to find an intensity for $b$ such that this transition takes place at an energy larger than the energy scale of our physical processes, so that as far as our gauge theory is concerned, this transition is not observed.

\begin{figure}[!ht]
  \includegraphics[width=\linewidth]{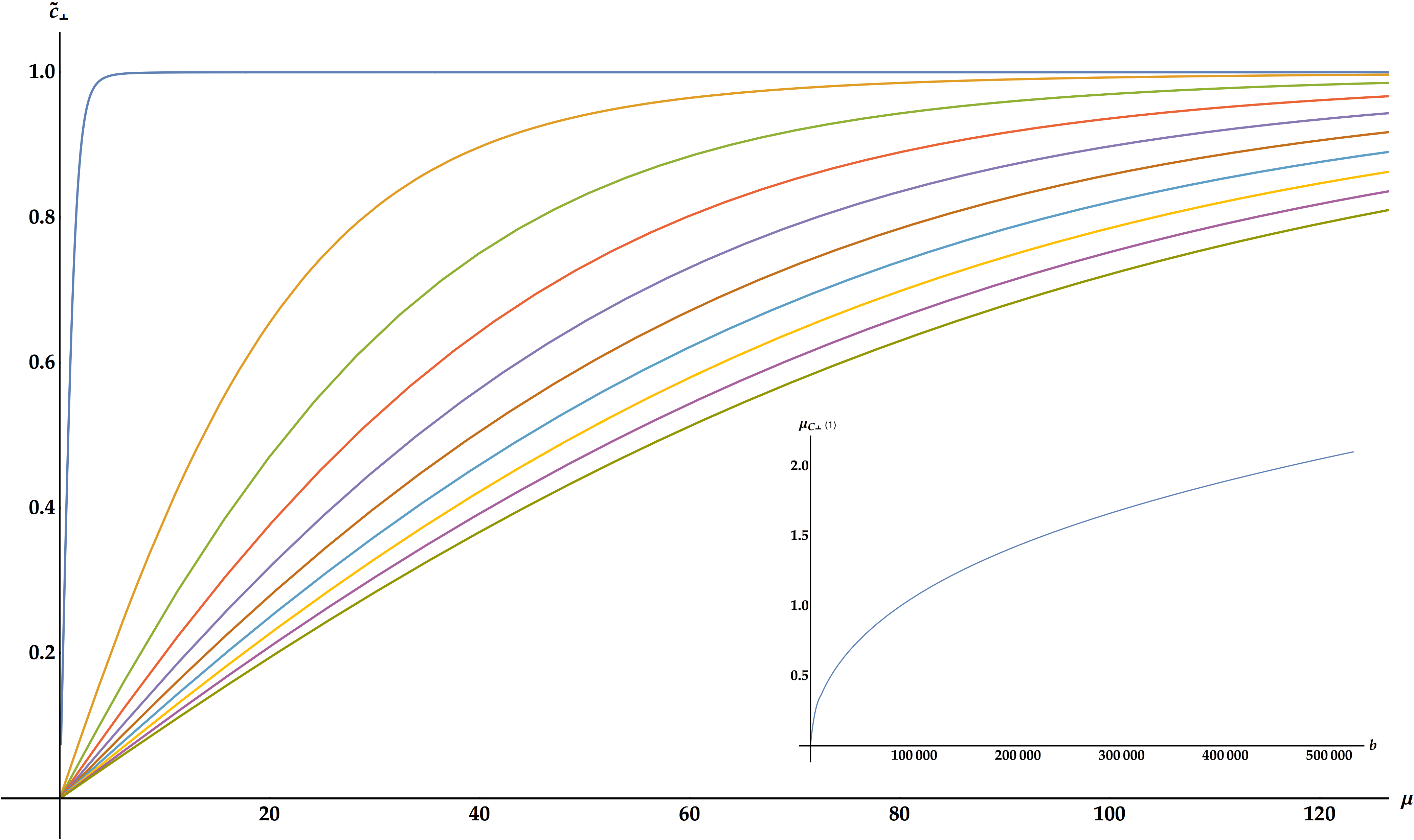}
\caption{$\tilde{c}_\perp$ as a function of the energy scale $\mu$ for $b=\{0, 4180,
10415, 17772, 25967, 34849, 44319, 54308, 64763, 75644\}$. The highest line corresponds to the $b=0$ case and increasing values of $b$ show plots that indicate the possibility of making $\tilde{c}_\perp$ as small as desired for any energy scale. The inset exhibits how the value of the energy scale below which $\tilde{c}_\perp$ is smaller than 1\% of the speed of light changes with $b$, and indicates that this energy scale grows with it, so $\tilde{c}_\perp$ can be kept as small as desired up to arbitrarily high energy scales by making $b$ more intense.}\label{lightcone1}
\end{figure}

\begin{figure}[!ht]
  \includegraphics[width=\linewidth]{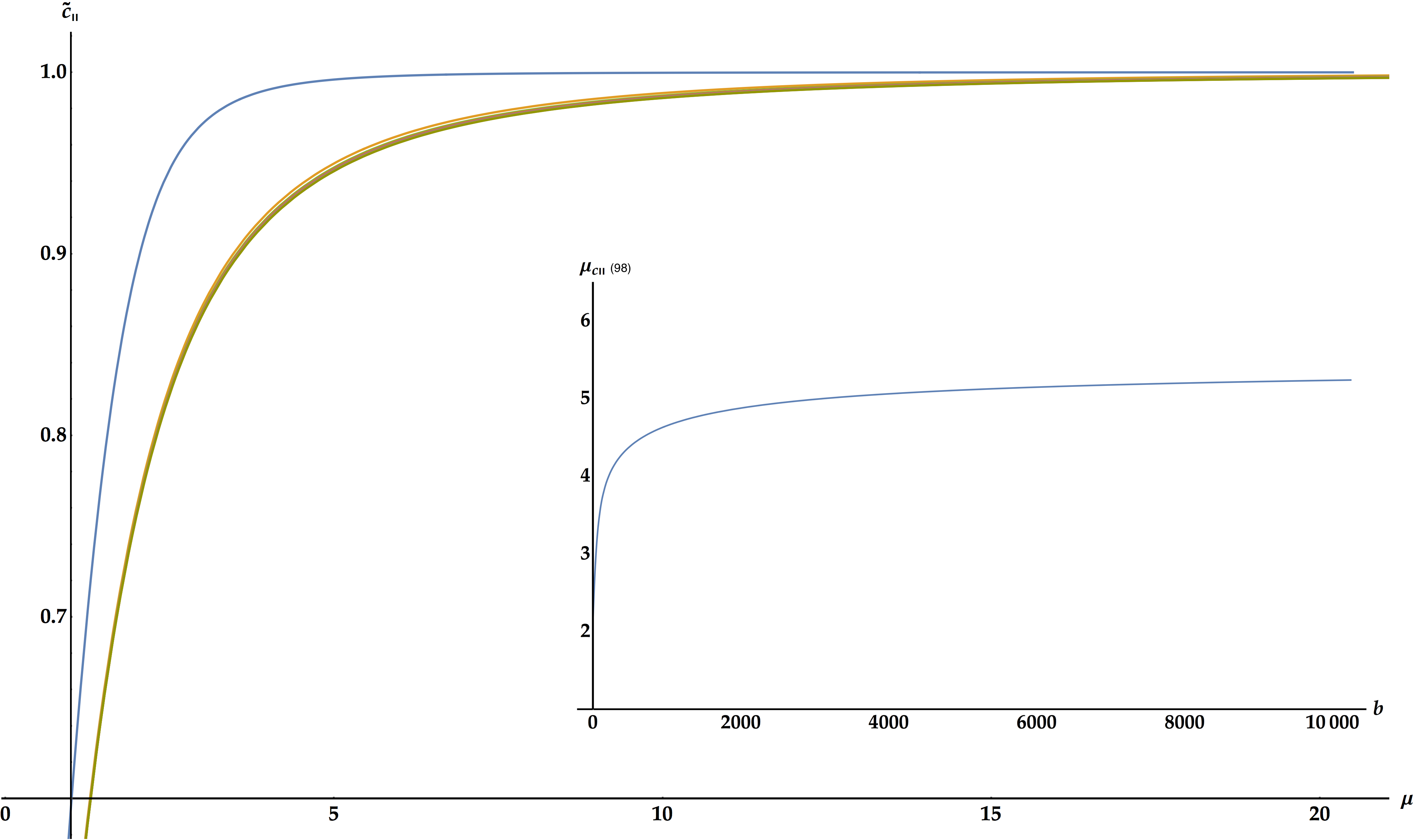}
\caption{$\tilde{c}_\parallel$ as a function of the energy scale $\mu$ for $b=\{0, 4180, 10415, 17772, 25967, 34849, 44319, 54308, 64763, 75644\}$. The highest line again corresponds to the $b=0$ case and increasing values of $b$ show plots that quickly converge to a given profile. The inset exhibits how the value of the energy scale below which $\tilde{c}_\parallel$ is smaller than 98\% of the speed of light changes with $b$, and indicates that this energy scale quickly approaches a constant as $b$ grows, so $\tilde{c}_\parallel$ gets close to the speed of light for low energy scales regardless of the intensity of $b$.}\label{lightcone2}
\end{figure}

In the plot (\ref{lightcone1}) we see that by increasing $b$, the limiting velocity $\tilde{c}_\perp=\sqrt{\frac{-g_{00}(r)}{g_{xx}(r)}}=\sqrt{\frac{-g_{00}(r)}{g_{yy}(r)}}=\sqrt{U/V}$ can be kept very small for any energy scale $\mu$ by increasing the intensity of $b$. On the other hand, the plot (\ref{lightcone2}) shows that, above a certain energy scale, the limiting velocity $\tilde{c}_\parallel=\sqrt{\frac{-g_{00}(r)}{g_{zz}(r)}}=\sqrt{U/W}$ cannot be pushed significantly away from 1 regardless of how intense the magnetic field is made.

To provide a different perspective and further understand how this effect compares in different directions, we define $\mu_{ci}(s)$ to be the energy scale up to which the limiting velocity in the $i$ direction remains smaller than $s\%$ the speed of light. The inset in (\ref{lightcone1}) shows how $\mu_{c\perp}(1)$ as an example can be made arbitrarily large by increasing the intensity of the field. On the other hand, the inset in (\ref{lightcone2}) shows for instance that $\mu_{c\parallel}(98)$ cannot be made higher than a certain value by intensifying the background field.

The results mentioned in the previous paragraphs provide evidence that propagation in the gauge theory is favored in the direction of the background field with respect to those perpendicular to it, consistent with a dimensional reduction taking place. Furthermore, these results exhibit that the dimensional reduction is effective up to an energy scale that grows with the intensity of the background magnetic field, reflecting the way in which the causal structure transitions from the infrared to the ultraviolet at higher energy scales for more intense magnetic fields.

\section{Conclusions}
\label{sec:conclusions}

By implementing the holographic version of the Wilsonian approach to renormalization we were able to determine properties of the corresponding group flow of a thermal gauge theory in the presence of a strong constant magnetic field. In particular we found the beta function for the coupling $\xi$ of the double trace fermionic operator $\bar{\Psi}(k)\Psi(k)$, and depicted our results in plots (\ref{fir}) to (\ref{guv}), where we see that, as in the zero temperature and no magnetic field case \cite{Laia:2011wf}, the renormalization flow happens between two fixed points, one in the low energy limit and the other in the high energy one.

The effect of the background magnetic filed is quite relevant and we can extract at least three ways in which it affects the behavior as the energy scales changes and that have direct impact, for instance, on the analysis of observational data.

The first conclusion we can draw is that the separation of the theories in the energy scale is increased by the introduction of the background magnetic field and grows with its intensity. We can see this in plots (\ref{fir}) and (\ref{gir}), where physics are fixed in the infrared and we observe that the transition to the ultraviolet theory happens at an energy scale that grows with the intensity of the magnetic field. This is consistent with what is reported in figures (\ref{fuv}) and (\ref{guv}), where physics are fixed in the ultraviolet and the transition to the infrared theory happens at lower energy scales as the intensity of the background field is increased. Even though the particulars of the flow of the coupling constant are scheme dependent, the generalities of it are robust, as argued in the main of the text.

The transcendence of this effect is that if when performing a high energy collision experiment we are interested in exploring the physics of a theory that happens as an ultraviolet limit of some renormalization flow, the energy that will be necessary to inject into the system to access the relevant processes will increase if a very intense magnetic field is present.

An example of how this observation can be relevant is that in a system like the quark gluon plasma obtained in experiments like RHIC or LHC, measurements taken from events with different centralities cannot be assumed to explore physics of the theory in the same energy scale, since the magnetic field intrinsic to the collision depends on how central the collision is. The events that will provide access to the ultraviolet physics at the lowest energy possible would be those coming from central collisions.

A theoretical exploration of this effect was given in section \ref{sec:DimRed}, where we saw that an intense enough magnetic field would make the four dimensional theory inaccessible for a large range of energies and would leave a dimensionally reduced effective theory.

The second conclusion, from plot (\ref{coup}), is that the difference in the coupling constants in the infrared and ultraviolet theories increases with the intensity of the field. So if fundamental physics are fixed at a very high energy scale, the apparent coupling that will be observed in a low energy experiment will depend on whether or not a magnetic field is affecting the theory. Given that this comparison is made at the fixed point of the theory, it is scheme independent.

The third effect that we want to comment on comes from the difference in the probabilities for propagation in directions parallel and perpendicular to the magnetic field. This difference would imply that the detected ellipticity for a collision would receive an extra contribution from the non centrality through this mechanism, making it larger than anticipated if this is not taken into account. This could be of particular relevance for experiments where measurements are used to determine the Fourier component v2 of the azimuthal anisotropy. Again, the anisotropy would prevail to higher energy scales for larger background magnetic fields in a scheme independent manner.

We think our study provides three ways in which the presence of a magnetic field can pragmatically affect data analysis in high energy physics.

\acknowledgments

We would like to thank Alberto G\"uijosa for helpful discussion, particularly concerning the subluminal limiting speed. LP highly appreciates the hospitality of the Physics Department of UCSB while working on the revised version of this work, and in particular thanks David Berenstein and Mark Srednicki for helpful input about the scheme dependence of the results. We also acknowledge partial financial support from PAPIIT IN113115, UNAM.

\end{document}